\documentclass[onecolumn,aps,amsmath,amssymb]{revtex4}

\usepackage{graphicx}
\usepackage{dcolumn}
\usepackage{bm}

\newcommand{\be}{\begin{eqnarray}}
\newcommand{\ee}{\end{eqnarray}}

\begin{document}
\title{Induced Universal Properties and Deconfinement}

\author{\'Agnes {\sc M\'ocsy}}
 \email{mocsy@alf.nbi.dk}
 \author{Francesco {\sc Sannino}}
 \email{francesco.sannino@nbi.dk}
 \author{Kimmo {\sc Tuominen}}\email{tuominen@nordita.dk}
 \affiliation{The Niels Bohr Institute \& NORDITA, Blegdamsvej 17,
 DK-2100 Copenhagen \O, Denmark }
\date{May 2003}

\begin{abstract}
We propose a general strategy to determine universal properties
induced by a nearby phase transition on a non-order parameter
field. We use a general renormalizable Lagrangian, which contains
the order parameter and a non-order parameter field, and respects
all the symmetries present. We investigate the case in which the
order parameter field depends only on space coordinates, and the
case in which this field is also time dependent. We find that the
spatial correlator of the non-order parameter field, in both
cases, is infrared dominated and can be used to determine
properties of the phase transition. We predict a universal
behavior for the screening mass of a generic singlet field, and
show how to extract relevant information from such a quantity. We
also demonstrate that the pole mass of the non-order parameter
field is not infrared sensitive. Our results can be applied to any
continuous phase transition. As an example, we consider the
deconfining transition in pure Yang-Mills theory, and show that
our findings are supported by lattice data. Our analysis suggests
that monitoring the spatial correlators of different hadron
species, more specifically the derivatives of these, provides an
efficient and sufficient way to experimentally uncover the
deconfining phase transition and its features.
\end{abstract}

\maketitle
\section{Introduction}
\label{yksi}

The relevant properties of phase transitions are best investigated
using order parameters. However, sometimes it is profitable to
determine new universal features associated with non-order
parameter fields \cite{Sannino:2002wb,{Mocsy:2003tr}}. Besides, in
nature most fields are non-order parameter ones. To derive the
general properties of a generic singlet field we couple this to
the order parameter \cite{Mocsy:2003tr}. In this way we can
directly study the transfer of information, envisioned in
\cite{Sannino:2002wb}, of the phase transition properties from the
order parameter to the singlet field.

We analyze two cases: time dependent order parameter fields and
time independent ones. The non-order parameter field, always time
dependent, is a scalar under Lorentz transformations. Also, it is
a singlet under any symmetry group, in particular under the
symmetry group whose breaking is monitored by the order parameter.
We further assume the non-order parameter field to have a large
mass with respect to all the other scales in play, and hence, an
associated small correlation length near the phase transition.
Although time independent order parameter fields carry the
physical information about the phase transition, they do not
propagate and cannot be canonically quantized. An example is the
Polyakov loop, which is considered to be the order parameter of
the pure Yang-Mills theory, and by construction is a function only
of space. A generic singlet field in the Yang-Mills theory is the
glueball, which is also a physical state of the theory.

Time dependent order parameter fields are generally associated
with physical states. These can be either composite, such as the
chiral condensate, the order parameter related to chiral symmetry
in QCD, or elementary, such as the Higgs field might be for the
electroweak phase transition. Using general field theoretical
arguments we demonstrate that the screening mass associated with
the spatial two-point function of the singlet field is heavily
affected by the nearby phase transition for both time dependent
and time independent order parameters. The important fact is that
we can predict the general behavior of the static two-point function of
the singlet field, and associate it uniquely with the specific
character of the phase transition. More specifically, the
screening mass at one loop has a drop near the phase transition.
This is due to the three dimensional nature of the screening mass,
which makes it particularly sensitive to infrared physics, and it
explains the transfer of information between the order parameter
and the singlet field. When going beyond the one loop
approximation the drop of the screening mass of the scalar singlet
field, both for time independent and time dependent order
parameters, is finite at the phase transition. Furthermore, the
drop itself is controlled by the ratio of the square of the
relevant coupling of the singlet field to the order parameter and
the coupling governing the self-interaction of the order
parameter. The result for the three dimensional order parameter
has already been presented in \cite{Mocsy:2003tr}. Here we provide
details of the basic computations used in \cite{Mocsy:2003tr},
especially for the higher order corrections, as well as a proof of
the result for the screening mass in the time dependent case. For
the latter we also present the effects of the nearby phase
transition on the pole mass of the singlet field. We demonstrate,
somewhat surprisingly, that although we still observe a small
drop, this physical quantity is not infrared dominated. So in
general, only the spatial correlation lengths feel the presence of
a nearby phase transition.

The paper is organized as follows: In section \ref{kaksi} we
introduce the theory which is investigated in detail in section
\ref{kolme} for the case of an order parameter field that depends
only on the space coordinates, and in section \ref{nelja} for an
order parameter field that depends both on time and space. The
information conveyed in section \ref{kolme} complements and
enlarges the one presented in ref.~\cite{Mocsy:2003tr}. In section
\ref{viisi} we conclude.

Our main conclusion in both cases, space and space-time dependent
order parameter fields, is that the information about the phase
transition, encoded in the behavior of the order parameter field
is transferred to, and obtainable from the singlet field(s)
present in the theory. The present analysis suggests that
monitoring the spatial correlators even of heavy hadrons provides
an efficient and sufficient experimental way to uncover the
existence and features of the chiral/deconfining phase transition.
So if our results are phenomenologically applicable in
relativistic heavy ion collisions at RHIC, they would lead to a
clear signal for the existence of a deconfined phase at the early
instant of such a collision.

\section{General Set Up}
\label{kaksi}

We consider a temperature driven phase transition and work in a
regime close to the phase transition. In order for our results to
be as universal as possible, we use a renormalizable Lagrangian
containing a field neutral under the global symmetries, and the
order parameter field, charged under the symmetries, 
as well as their interactions. The
protagonists of our theory are two real fields, $h$ and $\chi$.
The field $h$ is a scalar singlet, while $\chi$ transforms
according to $\chi\rightarrow z\,\chi$ with $z \in Z_N$.

\begin{figure}[t]
\includegraphics[width=6truecm]{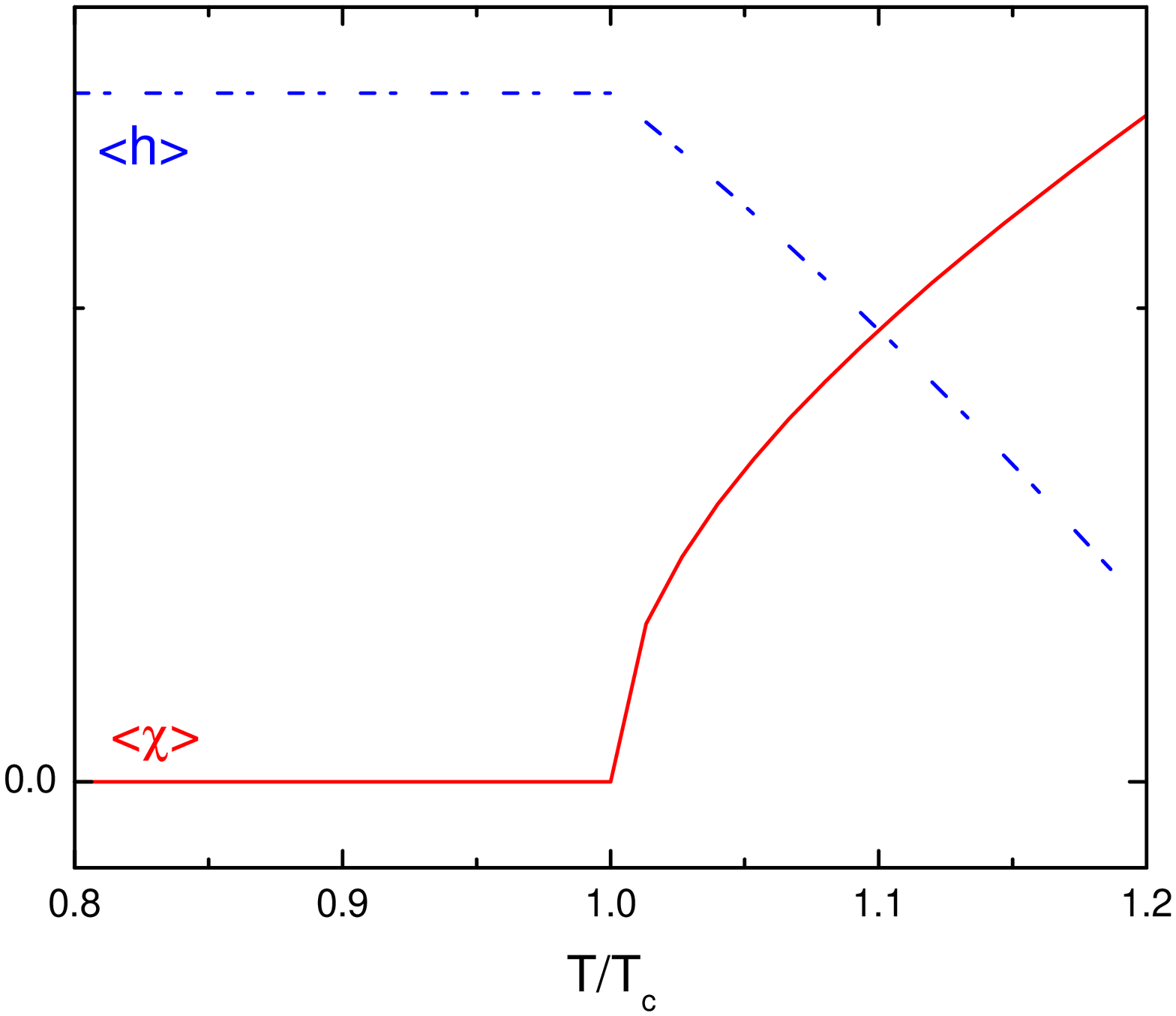}
\caption{Schematic behavior of the expectation values of the order parameter
$\langle \chi \rangle$ (solid--line) and the non-order parameter
$\langle h \rangle$ (dashed--line) fields close to the phase
transition as a function of the temperature. Note that $\langle h
\rangle$ is never zero and the constancy in the unbroken phase is
due to the tree level approximation.} \label{Figura}
\end{figure}

While the generalization to $Z_N$ is straightforward, we consider
explicitly the case of $Z_2$, which is suitable for understanding
the deconfining phase transition of two color Yang-Mills. This has
been heavily studied via lattice simulations
\cite{Damgaard,{Hands:2001jn}}. The most general renormalizable
potential is:
\begin{eqnarray}
V(h,\chi)&=&\frac{m^2}{2} h^2   + \frac{m^2_{0\chi}}{2}\, \chi^2 +
\frac{\lambda}{4!}\chi^4 + g_0 h + \frac{g_{1}}{2}\,h\chi^2 +
\frac{g_2}{4}\,h^2\chi^2 + \frac{g_3}{3!}h^3 + \frac{g_4}{4!}h^4
\, .\label{potential}
\end{eqnarray}
The coefficients are all real and stability requires $\lambda \geq
0$ and $g_4 \geq 0$. We further assume $g_1>0$ and $g_0<0$. At
this level we did not yet commit on the space-time dependence of
the fields, although we have normalized them as if they were
living in $3+1$ dimensions. This renormalizable potential can be
considered, for example, as a truncation of the one presented in
\cite{Sannino:2002wb}, and can be used to determine some of the
space-time independent properties of the vacuum. In order to go
beyond constant field approximation, we must specify the kinetic
terms. Since the singlet field $h$ is physical, it has the
ordinary standard four dimensional kinetic term,
i.e.~${\partial_{\mu}h\partial^{\mu}h}$. For the $\chi$ field we
first consider the case in which $\chi$ is time independent. In
this case the associated kinetic term reads as ${\nabla\chi
\nabla\chi}$. We will then consider the case in which $\chi$
experiences also time dimension, and thus the kinetic term is
${\partial_{\mu}\chi\partial^{\mu}\chi}$. The time and space
independent order parameter field, and its interactions with a
singlet field were introduced in \cite{Sannino:2002wb}, in order
to understand the transfer of information from Polyakov loops to
glueballs in pure Yang-Mills theories. The first studies of a
renormalizable version of the previous idea, in which the order
parameter is a function only of the space dimensions, were
performed in \cite{Mocsy:2003tr}. Here we review and extend the
analysis of \cite{Mocsy:2003tr}, and show the fundamental
differences and similarities with respect to a time dependent
order parameter field. Before proceeding we state the assumptions
under which we conduct our analysis for the time (in)dependent
order parameter: i) The $\chi$ field is light close to the
transition, hence it dominates the dynamics. ii) The $h$ field is
heavy, and thus we can neglect its quantum and Boltzman suppressed
thermal loop corrections. The extremum of the linearized potential 
(in the $h$ field) is for
\begin{eqnarray}
\langle\chi\rangle^2 &=& -6 \frac{m^2_{\chi}}{\lambda} \qquad {\rm
with} \quad m^2_{\chi} \simeq m^2_{0\chi} + g_1\langle h \rangle \
,
\quad {\rm and } \quad \langle h \rangle\,\, \simeq -
\frac{g_0}{m^2}-\frac{g_1}{2m^2}\langle\chi\rangle^2 \, .
\label{vevs}
\end{eqnarray}
Here $m_{\chi}$ vanishes at the phase transition.
%
%
Having $g_1>0$ and $g_0<0$, together with $3g_1^2<\lambda m^2$,
assures the positivity of these expectation values, and with this
choice also the extremum of the potential is a minimum. A more
complete treatment would require to go beyond the linearized
approximation. The temperature dependence of $\langle\chi\rangle$
and $\langle h\rangle$ is qualitatively sketched in figure
\ref{Figura}. Note, that in (\ref{potential}) the term linear in
$h$ is removed, at tree level, by shifting the field as follows:
\begin{eqnarray}
h \rightarrow \langle h\rangle + h \ .
\label{shift}
\end{eqnarray}
Such term, however, is regenerated at higher orders, correcting
the $h$ expectation value. The effect of the shift leads to
modified coefficients in the Lagrangian.  For example:
\begin{eqnarray}
g_0 &\rightarrow &0 \ ,\qquad g_1\rightarrow g_1 + g_2\,\langle h
\rangle \ , \qquad m^2_{0\chi} \rightarrow  m^2_{0\chi} +
g_1\,\langle h\rangle + \frac{ g_2}{2} \,\langle h\rangle^2 \ .
\label{modcouplings}
\end{eqnarray}
In the following we consider the fluctuations of $h$ around its
vacuum expectation value.
%

\section{Time-Independent Order Parameter Field~: Probing Static Properties}
\label{kolme}

The first difference between a purely spatial order parameter and
a time dependent one is that in the first case no Bose-Einstein
distribution emerges for $\chi$ in thermal equilibrium. More
specifically, we are allowed by construction to consider only
spatial fluctuations. Besides, since we assumed the field $h$ to
be heavy compared to the relevant scales at the phase transition,
including the temperature, the $h$ induced thermal corrections can
be safely neglected. We postulate, as customary, that $m_{\chi}$
is a function of temperature and vanishing at the transition
point. Interestingly, since the order parameter field is purely
three dimensional, we can freely choose from which direction the
symmetry is restored. Higher order temperature corrections cannot
reverse the direction of symmetry restoration. As we discuss in
detail in the next section, this is not the case if the order
parameter field is time dependent. Having in mind the Yang-Mills
deconfining phase transition, we assume that the $Z_2$ symmetry is
restored at low temperatures. Our analysis is not affected when
reversing the direction of the transition.

We have chosen $\chi=\chi({\bf{x}})$, and decomposed the four
dimensional field $h$ into its Matsubara modes. After integrating
over time the action reduces to an effective three dimensional
one. In the three dimensional case the kinetic term for $h$ and
its self-interaction terms receive contributions from all
Matsubara modes. However, only the zero mode contributes to the
$h\chi^2$-interaction, which is the one driving the dynamics of
the $h$ field close to the phase transition \cite{Mocsy:2003tr},
as the $\chi$ field becomes light with respect to $h$. Hence, we
confine our discussion here to the theory which features the
fields $\chi$ and $h_0$, and study the spatial fluctuations. For
simplicity, in this section we denote $h_0$ by $h$ which is taken
to be directly the fluctuation field around its tree level vacuum
expectation value. The three dimensional Lagrangian reads:
\begin{eqnarray}
-{\cal L}_3=\frac{1}{2}\nabla h\nabla
h+\frac{1}{2}\nabla\chi\nabla\chi+ \frac{1}{2}m^2h^2
+\frac{1}{2}m^2_{\chi}\chi^2\ + T\frac{\lambda}{4!}(\chi^2)^2 +
 \sqrt T \frac{g_{1}}{2}\,h\chi^2 +
T\frac{g_2}{4}\,h^2\chi^2 + \sqrt T\frac{g_3}{3!}h^3 +
T\frac{g_4}{4!}h^4 \, ,\nonumber\\
\end{eqnarray}
where the coupling constants have the same mass dimension of the
corresponding four dimensional theory. All of these are real with
$\lambda\ge 0$ and $g_4\ge 0~$.

Using the renormalizable theory defined by the above Lagrangian,
we now compute all one--loop corrections. We find the following
contributions to the $h$ one point function:
\begin{eqnarray}
\parbox{15mm}{\includegraphics[width=1.5cm,clip=true]{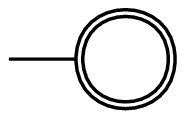}} +
\parbox{15mm}{\includegraphics[width=1.5cm,clip=true]{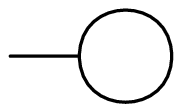}}
=\frac{\sqrt{T}}{8\pi }(g_1 m_\chi+g_3m),
\end{eqnarray}
which leads to temperature dependent corrections to the
expectation value of $\langle h \rangle$. We turn our
attention to the $h$ two point function. The full expression at
one-loop level is given by the following set of diagrams:
\be
\includegraphics[width=8.5cm,clip=true]{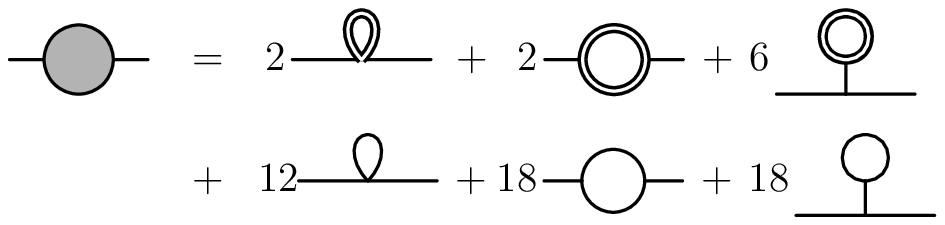}
\label{selfh} \ee Double lines indicate $\chi$ fields, while
single lines stand for the $h$ field. The number in front of each
diagram is the associated combinatorial factor. We regularize the
ultraviolet divergent contribution by subtracting the divergent
terms. The second diagram is infrared divergent for $T\rightarrow
T_c$ and gives the dominant one loop contribution to the screening
mass of $h$. In the limit of zero external momentum:
\be
\parbox{15mm}{\includegraphics[width=15mm,clip=true]{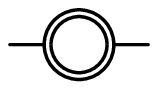}}&=&
T\,(\frac{g_1}{2})^2\int\frac{d^3k}{(2\pi)^3}\frac{1}{(k^2+m_\chi^2)^2}
=   T\, \frac{g^2_1}{32\pi m_{\chi}} \, . \label{one-loop}
\end{eqnarray}
It is worth mentioning that $\langle h(x)h(0)\rangle$ is
associated to the correlator $\langle\chi^2(x)\chi^2(0)\rangle$
when the couplings $g_2$, $g_3$ and $g_4$ are neglected since the
term $g_1h\chi^2$ is a source for $\chi^2$ when the $h$ kinetic
term is omitted.

After having taken proper care of the ultraviolet divergent terms
the finite contributions from the diagrams in equation
(\ref{selfh}) are
\be m^2(T)&=&m^2-T\,\frac{g^2_1}{16\pi  m_{\chi}} -
\frac{T}{8\pi}\,m_\chi\left({g_2} - \frac{g_1g_3}{m^2}\right)
-\frac{T}{16\pi}\,m\left(2{g_4} - \frac{g_3^2}{m^2}\right)\ ,
\label{bare3d} \ee
which clearly shows that the terms in parenthesis (third and
fourth terms) are subleading with respect to the second one near
the transition point. Equation (\ref{bare3d}) illustrates how the
nearby phase transition is directly felt by the non-order
parameter field, and furthermore, gives the general prediction
that the screening mass of the singlet field must {\it decrease}
close to the phase transition. If we stop the analysis here, at
one loop level, we predict the following critical behavior for
$\Delta m^2=m^2(T)-m^2$, where $m$ is the mass at a temperature
close to the critical point:
\begin{eqnarray}
\Delta m^2(T)&=& - \frac{g_1^2\,T}{16\pi\,m_{\chi}}
 \sim t^{-\frac{\nu}{2}} , \quad T<T_{\rm{c}}  \\
 \Delta m^2(T)&=& -
\frac{g_1^2\,T}{16\pi\,M_{\chi}}
 \sim t^{-\frac{\nu}{2}} ,  \quad  T>T_{\rm{c}}
\end{eqnarray}
where $M_{\chi}=\sqrt 2|m_\chi|\propto |T-T_c|^{\nu/2}$. This one
loop result breaks down at the transition point, due to the
infrared singularity. However, since $h$ is not the order
parameter field, its correlation length (i.e.~$1/m$) is not
expected to diverge at the phase transition. In the following we 
investigate the behavior of the $h$ screening mass near the phase 
transition by going beyond the one-loop approximation.

\subsection{Healing the IR behavior}

When analyzing contributions beyond one loop order to the $h$
two-point function, the number of diagrams, and distinct
topologies one needs to consider proliferates. We select a
subclass of diagrams, that heal the infrared divergences, while
capturing the essential physical properties of the problem at
hand. Besides, this subclass of diagrams we consider is known to
be exact in the large $N$ limit of a theory with $O(N)$ symmetry
\cite{Coleman:jh}, and the summation can be performed exactly. We
note that a large $N$ approximation might not be the best choice
for investigating the details of the phase transition, however we
show that it well reproduces the behavior of the screening mass of
hadronic degrees of freedom near the deconfinement phase
transition as given by lattice simulations. A finite result for
the two-point function, both in the broken and unbroken phases
will emerge, while the expression for the mass of $h$ turns out to
be continuous across $T_c$.

In the unbroken phase (i.e.~$T<T_c$) it is known, that for a
generic $O(N)$ theory in the large $N$ limit the following chain
of bubble diagrams represents the leading contribution,
\begin{eqnarray}
\includegraphics[width=6.6cm,clip=true]{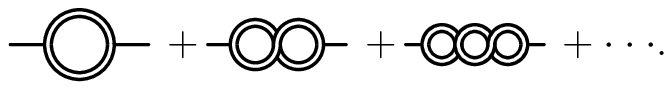} \nonumber
\end{eqnarray}
These diagrams constitute a geometrical series. For the $O(N)$
theory this resummation procedure is exact in the large $N$ limit
\cite{Coleman:jh}. We choose the same set of diagrams in the
unbroken phase. Denoting the dimensionless single loop integral by
${\cal{I}}$, the expression for a single diagram in the chain made
of $p$ loops, and $p-1$ vertices of $\chi^4$-type is:
\begin{eqnarray}
\frac{g_1^2}{2}(-\frac{\lambda}{2})^{p-1}{\cal{I}}^p \, .
\end{eqnarray}
The exact sum of all the diagrams in the chain is simply:
\begin{eqnarray}
g_1^2{\cal{I}}\frac{\frac{1}{2}}{1+\frac{\lambda{\cal{I}}}{2}} \,
\, , \qquad {\rm with } \qquad  {\cal{I}}=\frac{T}{8\pi m_\chi} \,
\, . \label{sum1}
 \end{eqnarray}
Using eq.~(\ref{sum1}) the following expression for $m$ follows:
\begin{eqnarray}
m^2(T)=m^2 - T\, \frac{g_1^2}{16\pi\,m_{\chi}+\lambda\,T} \, ,
\end{eqnarray}
which is finite at $T_c~$, where it yields:
\begin{eqnarray}
m^2(T_c)=m^2-\frac{g_1^2}{\lambda} \, .
\end{eqnarray}
This is the main result of \cite{Mocsy:2003tr}, predicting that
close to the phase transition the singlet state must have a
decreasing mass parameter, associated with spatial correlations.
More specifically, the drop at the phase transition point is given
by the ratio of the square of the coupling constant governing the
interaction of the singlet state with the order parameter
(i.e.~$g_1$) to the order parameter field self-interaction
coupling constant $\lambda$. In this way, via the drop of the
singlet field at the phase transition, one can derive a great deal
of information about the phase transition, and about the order
parameter itself.

The analysis is much more complicated in the broken phase. Indeed,
when $T>T_c$ $\chi$ develops an average value
$\langle\chi\rangle=v$, which induces one also for $h$ as shown in
eq.(\ref{vevs}). This is in agreement with the results found in
\cite{Sannino:2002wb}. So on general grounds, the fields $\chi$
and $h$ mix in this phase. Defining the mass eigenstate fields as
$H$ and $\Xi$, these are related to $h$ and $\chi$ via:
\begin{equation}
\left(%
\begin{array}{c}
  h \\
  \chi \\
\end{array}%
\right) =
\left(%
\begin{array}{cc}
  \cos\theta & -\sin\theta \\
  \sin\theta & \cos\theta \\
\end{array}%
\right)
\left(%
\begin{array}{c}
  H \\
  \Xi \\
\end{array}%
\right) \, .
\end{equation}
The mixing angle $\theta$ is proportional to $g_1\,v/m^2~$, and
therefore, the mixing can be neglected within the present
approximations, simplifying the analysis considerably. Like for
the symmetric phase, we consider only the effects due to the
$\chi$ loops for the $h$ propagator. Due to symmetry breaking we
now need to consider also the trilinear $\chi$ coupling
\begin{eqnarray} -\frac{\lambda}{3!}\,v\,\chi^3 \, ,\end{eqnarray}
which is expected to affect the analysis. At one loop level the
only diagram to compute is again the one in eq.~(\ref{one-loop}),
with $m_{\chi}$ replaced by $M_{\chi}$. Hence, on general grounds,
we predict a drop in the mass of $h$ also on the right hand side
of $T_c~$. In this phase the infrared divergence is still present
at one loop level. Curing such divergence is now more involved due
to symmetry breaking. In the case of the large $N$ limit of $O(N)$
symmetry, one can show, that diagrams with trilinear vertices are
again suppressed relative to the simple bubble diagrams.

Here we go beyond the large $N$ limit by computing a new set of
diagrams, which can be evaluated exactly, and thus capture
relevant corrections due to symmetry breaking, neglected in the
large $N$ limit. The new chain of diagrams we compute has terms of
the form:
\begin{equation}
\includegraphics[width=6.5cm,clip=true]{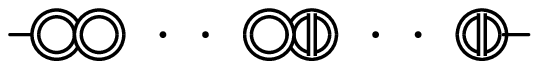}
\label{bubbles2}
\end{equation}
This class of diagrams has knowledge about the onset of the
symmetry breaking via the presence of the trilinear vertices, and,
in that respect, is the simplest extension of the chain of simple
bubble diagrams. Another amusing property of (\ref{bubbles2}) is,
that the sum can be performed exactly, as we now demonstrate.

Denoting the number of simple bubbles by $p$, and the number of
bubbles with trilinear vertices by $q$, the total number of loops
is $L=p+2q$, and the number of $\chi^4$ vertices is $p+q-1$, while
that of the trilinear vertices is $2q$. The loop integral of the
single bubble is ${\cal{I}}$. The loop integral of the bubble with
two trilinear vertices yields ${\cal{I}}^2/9$. The
generic single multi-loop diagram of type (\ref{bubbles2}), with
$p$ ordinary bubbles and $q$ bubbles with two trilinear vertices,
yields the following contribution:
\begin{eqnarray}
\frac{g_1^2}{4}2\left (-\frac{\lambda}{4!}\right )^{p+q-1}\left
(\frac{\lambda v}{3!}\right )^{2q}\left (\frac{3}{\lambda
v^2}\right )^q\frac{{\cal{I}}^{p+2q}}{3^{2q}} \left (3!\right
)^{2q}
\left (\frac{4!}{2}\right )^{p+q-1}\left(%
\begin{array}{c}
  p+q \\
  q \\
\end{array}%
\right)=\frac{g_1^2}{2}\left (-\frac{\lambda}{2}\right
)^{p+q-1}\left (\frac{\lambda}{3}\right )^q{\cal{I}}^{p+2q}
\left(\begin{array}{c}
  p+q \\

  q \\
\end{array}%
\right ) \, . \nonumber \\
\end{eqnarray}
On the left hand side of the equation the binomial factor
corresponds to the number of different ways to distribute $q$
rungs into $p+q$ loops, and the preceding factorials are the
associated combinatorial factors, due to the permutation of the
vertices.

The infinite sums over $q$ and $p$ are evaluated by replacing the
summation over $p$ with the summation over the total number of
loops $L$. For fixed $L$, $q$ can take values $q=0,1,\dots,\lfloor
L/2\rfloor$, where $\lfloor x\rfloor$ is the largest positive
integer smaller than or equal to $x$. Then the sum becomes
\begin{eqnarray}
\frac{g_1^2}{2}\frac{(-2)}{\lambda}\sum_{L=1}^{\infty}\left
(-\frac{\lambda{\cal{I}}}{2}\right )^L
\sum_{q=0}^{q_{\rm{max}}}\left (-\frac{2}{3}\right)^q \left
(\begin{array}{c}
  p+q \\
  q \\
\end{array}
\right)&=&\frac{g_1^2}{2}\left (-\frac{2}{\lambda}\right
)\sum_{L=1}^\infty\left (-\frac{\lambda{\cal{I}}}{2}
\sqrt{\frac{2}{3}}\right )^L\,U_L\left
(\frac{1}{2}\sqrt{\frac{3}{2}}\right )\, , \label{sum1/2}
\end{eqnarray}
where the function $U_L(x)$ appearing in the intermediate step is
the Chebychev polynomial of the second kind. Remarkably, the final
sum can be performed exactly, yielding for $T>T_c$ again an
infrared finite result:
\begin{eqnarray}
m^2(T)&=& m^2-\frac{g^2_1{\cal
I}}{2}\,\frac{1+\frac{\lambda}{3}{\cal
I}}{1+\frac{\lambda}{2}{\cal I}+ \frac{\lambda^2}{6}{\cal I}^2} \
, \\{\cal I}&=&\frac{T}{8\pi\,M_{\chi}}\, .
\end{eqnarray}

We see that $m(T_c)$ from the broken side of the transition equals
exactly the one from the unbroken side of the transition, even
when departing from the large $N$ limit. The mass squared of $h$
is a continuous function through the phase transition, and the
associated correlation length remains finite. This result does not
hold order by order in the loop expansion, but only when the
infinite sum of the diagrams is performed. Although admittedly,
there is ambiguity in selecting the class of diagrams to take into
account, guided by the large $N$ limit in the symmetric phase, we
considered only the bubble chain of diagrams. However, in the
broken phase we chose to resum a new and richer class of diagrams.
The new class contains not only the relevant large $N$
contribution, but also carries information related to the
spontaneously broken symmetry. This class, although more involved,
can be summed exactly.

In order to disentangle relevant properties of the phase
transition, we construct slope parameters for the
singlet field:
\begin{eqnarray}
    {\cal D}^\pm &\equiv & \lim_{T\rightarrow T_{\rm{c}}^\pm}
    \frac{1}{\Delta m^2(T)} \frac{d\, m^2(T)}{dT} \, ,
    \label{slopes1}
\end{eqnarray}
with $\Delta m^2(T_c)= g_1^2/\lambda~$. The functional form of
$\cal{D}^{+}$ and $\cal{D}^{-}$ is the same, provided that the
same class of diagrams is resummed on both sides of the
transition. In this case we have:
\begin{eqnarray}
{\cal D}^- = \frac{16\,\pi}{\lambda\,T_c}\lim_{T\rightarrow
T_c^-}\frac{d\, m_{\chi}}{dT}
\end{eqnarray}
in the symmetric phase, and to obtain ${\cal D}^+$ in the broken
phase is sufficient to replace $m_{\chi}$ with
$M_{\chi}=\sqrt{2}|m_\chi| $. While the mass of $h$ remains finite
at $T_c$, its slope encodes the critical behavior of the theory.
For example, if $m^2_{\chi}$ vanishes as $(T_c-T)^{\nu}$ close to
the phase transition (with the correlation length $\xi\propto |T -
T_c|^{-{\nu}/{2}}$), then ${\cal D}^\pm$ scales with exponent
$(\nu/2 -1)$. A difference in the functional form of the slopes
may emerge, when on the two sides of the transition different
class of diagrams are resummed.
Using the wider class of diagrams in the broken phase considered
above while retaining just the simple bubble sum in the unbroken
we determine:
\begin{eqnarray}
{\cal D}^+
\simeq -3\,\frac{16\pi}{\lambda T_c}{|m_{\chi}|}{\cal D}^- \, .
\end{eqnarray}
This is due to the onset of spontaneous symmetry breaking, i.e. in
the broken phase the class of resummed diagrams contains trilinear
type of interactions. We identify thus a less singular behavior
with respect to the simple sum of bubbles. More specifically, the
scaling exponent for ${\cal D}^+$ is now $(\nu -1)$. While the
explicit relations between the scaling exponents and the slopes
are only valid within the given summation scheme, these quantities
are, nevertheless, a measure of the critical behavior near the
phase transition. Only experimental results will be able to select
which class best describes the data.

\begin{figure}[]
 \includegraphics[width=12truecm, height=4.5truecm]{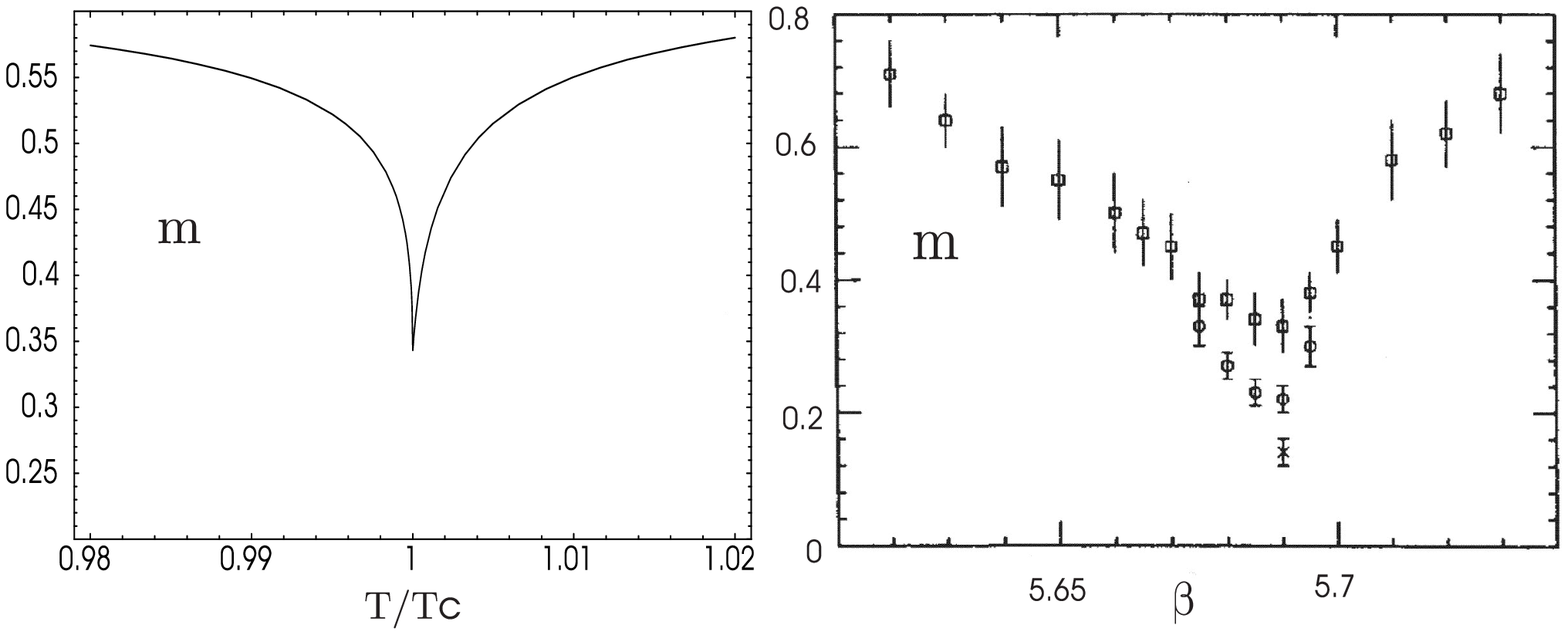}
\caption{Left panel: behavior of the mass of the singlet field
close to the phase transition as function of the temperature. 
Right panel: Lattice data; figure from \cite{Bacilieri:mj}.}
\label{Figura1}
\end{figure}

In figure \ref{Figura1} we schematically represent the behavior of
the $h$ screening mass as a function of the temperature, in units
of the critical temperature, for $m^2_{\chi}\propto (T_c-T)~$.
The left panel of fig. \ref{Figura1} illustrates the rapid
decrease of the singlet field screening mass in the critical
region. 
The right panel of fig. \ref{Figura1} is the
lattice data of \cite{Bacilieri:mj} obtained for the three color
Yang-Mills theory. The mass in \cite{Bacilieri:mj} is the glueball
screening mass. 
The resemblance between our results and the lattice results is
intriguing, but more lattice data is needed in order to
distinguish between the above possibilities, and quantitatively
determine the size of the drop.

It is common practice to isolate the order parameter field, and
more generally, the light degrees of freedom, since they are
expected to be the relevant degrees at low energies. While this is
certainly correct, we have explicitly shown how the essential
features of the phase transition are encoded in the non-order
parameter fields of the theory as well. Our analysis is useful
whenever the order parameter field is neither a physical quantity,
nor is phenomenologically accessible (e.g.~the Polyakov Loop).

\section{Time Dependent Order Parameter Field}
\label{nelja}

In this section we investigate the effects of the four dimensional
$\chi=\chi({\bf{x}},t)$ order parameter field on the properties of
a singlet $h=h({\bf{x}},t)$ field. Some of these effects turn out
to be significantly different than for the time independent case.
The order parameter field is now a physical field, and as such, it
can propagate in time, and can be canonically quantized. The
four-dimensional Lagrangian of the renormalizable theory we use to
define our Feynman rules is:
\be {\cal L}_4=\frac{1}{2}\partial_{\mu}h \partial^{\mu}h
+\frac{1}{2}{\partial_{\mu}} \chi {\partial^{\mu}} \chi -
\frac{m^2}{2} h^2 - \frac{m^2_{\chi}}{2}\, \chi^2 -
\frac{\lambda}{4!}\chi^4 - \frac{g_{1}}{2}\,h\chi^2 -
\frac{g_2}{4}\,h^2\chi^2 - \frac{g_3}{3!}h^3 - \frac{g_4}{4!}h^4
\, . \label{lagrangian4} \ee
All coupling constants are real, and $\lambda\ge 0$, $g_4\ge 0$,
and $g_2$ are dimensionless, while $g_1$ and $g_3$ have mass
dimension one. The largest scale is the mass $m$ of the $h$ field,
which, by our assumption, is again much larger than $m_\chi$, the
mass of the $\chi$ field, and it is also much larger than the
temperatures involved. We write then the couplings in terms of
dimensionless ones via $g_1=\hat{g}_1m~$, $g_2=\hat{g}_2~$, and
$g_3=\hat{g}_3m~$. The Lagrangian (\ref{lagrangian4}) contains all
the relevant and marginal operators: $g_1h\chi^2$ and $g_3h^3$ are
relevant operators, while, for example, $g_2h^2\chi^2$ is
marginal. We expect the relevant physics to be well described by
this set of operators.

The field $\chi$ is subject to $Z_2$ symmetry. Having the
Yang-Mills deconfining phase transition in mind, in the theory
with static order parameter field we {\it chose} the symmetric
phase to be at low temperatures, and thus the symmetry was broken
for temperatures greater than a phase transition temperature.
Accordingly, $m_\chi^2> 0$ for $T<T_c$ and $m_\chi^2< 0$ for
$T>T_c$. As we anticipated though in the previous section, when
the order parameter field is non-static we cannot simply choose
the direction of symmetry restoration a priori. In four-dimensions
a Bose-Einstein distribution emerges for $\chi$, and thus thermal
fluctuations become important. Thermal fluctuations have the
tendency to restore symmetry, destroying the possibility for the
formation and existence of any physical condensate at high
temperatures. The direction of the phase transition is therefore
important, and is completely determined by the theory. In the
following, we adopt the standard picture for the direction of the
phase transition, consistent with the effect of higher order
corrections to $m_\chi$: symmetry restoration sets in at high
temperatures \footnote{There are however exceptions as discussed
in \cite{Weinberg,Mohapatra}}. This is one of the major
differences between the theories with three- and four- dimensional
order parameter field.

\subsection{Diagramatics}

Based on Lagrangian (\ref{lagrangian4}) the diagrams contributing
to the $h$ two-point function at one-loop order are shown on
(\ref{selfh}). Even though these are the same as in the three
dimensional theory, here they are computed taking into account the
time dependence of $\chi$. We can show, that finite temperature
corrections from all diagrams that involve an $h$ loop are
Boltzman suppressed due to the heaviness of $h$ . Accordingly, the
relevant diagrams in (\ref{selfh}) for the singlet $h$ field are
those on the first line: the first (tadpole A), the second (eye),
and the third (tadpole B). Unlike in the three dimensional, time
independent case, where UV divergent tadpoles have been removed by
renormalization, in the four dimensional, time dependent case,
these diagrams also provide non-negligible temperature
contributions. Recall, that previously only the zero mode $h_0$
coupled in the $h\chi^2$-interaction. This is clearly not the
situation here: all modes of $h$, and all Matsubara modes of
$\chi$ participate in the interactions. The diagrams are evaluated
using standard techniques of the imaginary time formalism
\cite{kapusta}.
\be
\parbox{15mm}{\includegraphics[width=16mm,clip=true]{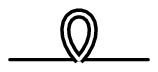}}&=&
-\frac{g_2}{4}T\sum_m\int\frac{d^3k}{(2\pi)^3}\frac{1}{\omega_m^2+\omega^2}
=-\frac{g_2}{4}\int\frac{d^3k}{(2\pi)^3}\frac{1}{2\omega}(1+2f(\omega/T))\,
,\label{tad} \ee
where $\omega=\sqrt{{\bf{k}}^2+{m_\chi}^2}$ is the energy, and
$\omega_m=2\pi mT$ is the Matsubara frequency of the internal
$\chi$-line. The second line in (\ref{tad}) is the result of the
frequency summation, and $f(x)=1/(e^x-1)$ is the Bose-Einstein 
distribution function. The term proportional to the number 1
before the distribution function represents the vacuum
contribution, and is divergent in the ultraviolet. This divergence
is absorbed in the mass renormalization process. The second term
contains no divergence, since it is regularized by the
distribution function. We focus on the physics in the phase
transition region, where $m_\chi\ll T$. Since the mass of the
order parameter field, $m_\chi~$, vanishes at $T_c$, we use the
high temperature limit to obtain an analytic result for the $h$
self-energy contribution
\be \Pi_{tadpole}^A \simeq\frac{g_2}{24}\,T^2\, .
\label{tadpoleA}\ee
Similarly, the other tadpole contributes with
\be \Pi_{tadpole}^B \simeq -\frac{g_1g_3}{24}\,\frac{T^2}{m^2} \,
.\label{tadpoleB}\ee
These contributions are real and provide temperature dependence to
the $h$ mass, not present in the three dimensional theory. In
terms of dimensionless couplings
\be \Pi_{tadpole}=
\left(\hat{g}_2-{\hat{g}_1\hat{g}_3}\right)\,\frac{T^2}{24}\, .
\label{tadh} \ee
Contribution to the self-energy of $h$ from the eye diagram is
\be
\parbox{15mm}{\includegraphics[width=15mm,clip=true]{fool.eps}}&=&
\left(\frac{g_1}{2}\right)^2T\sum_m\int\frac{d^3k}{(2\pi)^3}
\frac{1}{\omega_m^2+\omega^2}\frac{1}{(\omega_m-\omega_n)^2+\omega^2}
\nonumber\\
&=&\left(\frac{g_1}{2}\right)^2\int\frac{d^3k}{(2\pi)^3}\frac{1}{\omega}
\frac{1}{4 \omega^2+\omega_n^2}(1+2f(\omega/T)) \, .\ee
Here $\omega_n$ is the Matsubara frequency of the external line
and $\omega=\sqrt{{\bf{k}}^2+{m_\chi}^2}~$. We have restricted the
study to zero external momentum. In order to obtain physical
quantities from this real Euclidean integral we analytically
continue this to Minkowski space. After the replacement
$i\omega_n\rightarrow E+i\epsilon$ we obtain
\be \Pi_{eye}(E) =
-2\left(\frac{g_1}{2}\right)^2\int\frac{d^3k}{(2\pi)^3}\frac{1}{\omega}
\frac{1}{E^2-4 \omega^2}(1+2f(\omega/T)) \, .\label{eye}\ee

Another significant difference with respect to the theory with
time independent order parameter field is the existence of a real
and an imaginary part of the self-energy, that contributes to the
two-point function $\left[E^2-m^2-\Pi(E)\right]^{-1}~$. Here $m$
is the tree level mass of $h$ and at one-loop order
$\Pi=\Pi_{tadpole}+\Pi_{eye}~$. The real and imaginary parts can
be extracted from (\ref{eye}) by using
\be \frac{1}{z\pm i\epsilon}=P\frac{1}{z}\mp i\pi\delta(z)\, . \ee
Accordingly, the real part is given by the principal value of
(\ref{eye})
\be \mbox{Re}~\Pi_{eye}(E)
=-2\left(\frac{g_1}{2}\right)^2\frac{1}{2\pi^2}P\int dk
\frac{k^2}{\omega} \frac{1}{E^2-4 \omega^2}(1+2f(\omega/T))\, .
\label{eyereal}\ee
The term proportional to the number 1 before the distribution
function is again the vacuum contribution, divergent in the
ultraviolet, and is removed by usual renormalization. This real
part represents a shift in the mass squared of the $h$, and for
$h$ at rest, through its definition, determines the pole mass $M$:
\be {M}^2-m^2-\Pi_{tadpole}-\mbox{Re}~\Pi_{eye}(E=M)=0\, .
\label{polemass} \ee
Self-consistent numerical solutions of the above equation show
that the large tree level mass is dominant, and loop corrections
are negligibly small in the temperature range of interest,
i.e.~near the phase transition. Another interesting fact is, that
$M\simeq m$ acts as an infrared cutoff guaranteeing the absence of
infrared divergence for the pole mass. As a consequence, in this
case there is no need for the resummation of higher order
diagrams. An analytic expression in powers of $T^2/m^2$, and
setting $m_{\chi}=0~$, can thus be obtained in the region we work
in, namely $m_\chi\ll T\ll m$,
\be \mbox{Re}~\Pi_{eye}\left(E = m\right) =
-2\left(\frac{g_1}{2}\right)^2
\left(\frac{1}{6}\frac{T^2}{m^2}+\frac{4\pi^2}{15}\frac{T^4}{m^4}\right)
\simeq -\frac{\hat{g}_1^2}{12}m^2\frac{T^2}{m^2}\, . \ee
The absence of the infrared divergence in the pole mass is another
relevant difference between the two theories discussed in this
paper. This becomes apparent when summing all the one-loop
contributions from (\ref{tadh}) and (\ref{eyereal}):
\be M^2= m^2\left[1 +\left(\hat{g}_2-\hat{g}_1\hat{g}_3
-2\hat{g}^2_1\right)\,\frac{T^2}{24\,m^2} + {\cal
O}\left(\frac{T^4}{m^4} \right)\right]\, . \label{pole}\ee
The imaginary part of the self energy represents the net rate of
$h$-decay, and its evaluation provides no difficulties:
\be \mbox{Im}~\Pi_{eye}(E) = -2
\left(\frac{g_1}{2}\right)^2\frac{1}{16\pi}\Theta(E-2m_\chi)
\sqrt{1-\frac{4m_\chi^2}{E^2}}(1+2f(E/2T))\, . \label{imag} \ee
By rewriting $(1+2f)=(1+f)^2-f^2$ it is easy to understand that
above the threshold, $E=2m_\chi~$, (\ref{imag}) determines the
decay process $h\rightarrow\chi\chi~$, and the annihilation
$\chi\chi\rightarrow h~$, weighted by the thermal distribution,
with the net decay rate given by
\be \Gamma = -\frac{\mbox{Im}~\Pi_{eye}(E)}{2E} \, .\ee

Unlike in the three dimensional case, here there are finite
temperature corrections also to the two-point function of the now
dynamic order parameter field. The full set of one-loop diagrams
contributing to the $\chi$ field in the symmetric phase, together
with their corresponding combinatorial factors is:
\be
\includegraphics[width=13cm,clip=true]{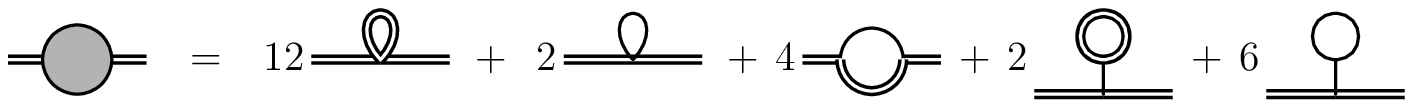}\label{chisym.fig}
\ee
Due to the existence of a cubic $\chi$ self-coupling in the broken
phase, there are, in addition to (\ref{chisym.fig}), two more
diagrams:
\be
\includegraphics[width=4.2cm,clip=true]{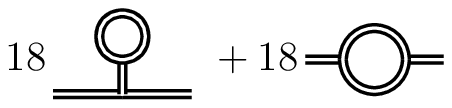}\label{chibrok.fig}
\ee
Here we are not going to discuss further details about the $\chi$
two-point function. The evaluation of these diagrams is standard.
There are two points we wish to stress though: First, that
contrary to the three dimensional theory, there are explicit
thermal fluctuations providing temperature dependence to the mass
of the order parameter field; Second, this mass is zero at the
critical temperature of the second order phase transition, and has
thus $m^2_\chi\propto (T_c-T)$ behavior.

\subsection{Static Properties}

In the three-dimensional theory we found an infrared divergent
result for the one-loop contribution to the $h$ two-point
function. This divergence came from what we call the eye diagram.
In four dimensions, as we have shown above, besides the eye
diagram there are two tadpoles also contributing. Here we show
that the static limit of the two-point $h$ function for the time
dependent order parameter field displays the same features as for
the time independent order parameter field case.

Here it is important to distinguish between the pole mass and the
screening mass. The pole mass, $M$, is defined through
(\ref{polemass}) as the pole in the full two-point function. The
screening mass, $m_s$, is defined by the location of the pole in
the static propagator for complex momentum $p=im_s$,
\be p^2+\Pi(E=0,p)=0 \, .\ee
In the small momentum limit this leads to the following definition
\be m_s^2=m^2+\lim_{p\rightarrow 0}\Pi(E=0,p)\, . \ee
In three dimensions the pole and screening masses are one and the
same thing, since the static propagator is the full propagator.

We have shown, that when looking at the pole mass the IR problem
of the eye diagram is regulated by the heavy $h$ mass. When
analyzing the screening mass, however, the one-loop IR divergence
present in the three dimensional case is recovered. Based on the
above definition, we set $E=0$ in expression (\ref{eyereal}). This
corresponds to effectively reducing the 4 dimensional theory to a
3 dimensional one. To display the relevant infrared contribution
we take the high temperature expansion in (\ref{eyereal}) which
yields:
\be 2\left(\frac{g_1}{2}\right)^2\frac{T}{4\pi^2}\int_0^\infty
dp\frac{p^2}{(p^2+m_\chi^2)^2} = g_1^2\frac{T}{32\pi m_\chi}\, .
\label{red4d} \ee
Hence, for the static limit the phase transition region is
dominated by the same type of infrared divergence we encountered
in the time independent order parameter field case. Note, however,
that the numerical constant in eq.\ (\ref{red4d}) differs from
that in the second term of eq. (\ref{bare3d}). The reason for this
is that the reduction, both here and in section \ref{kolme}, was
done only for the modes of $h$, and thus in eq. (\ref{red4d}),
contrary to (\ref{bare3d}), all Matsubara modes of $\chi$
contribute.

By combining all the diagrams we find for the screening mass at
one-loop order
\be m^2\left[1 - \frac{\hat{g}_1^2}{32\pi}\frac{T}{m_\chi} -
\left({\hat{g}_1\hat{g}_3} -
{\hat{g}_2}\right)\frac{T^2}{24\,m^2}\right]\, ,  \ee
showing clearly the eye contribution as the infrared dominant one.
Note, that above we have tree level coefficients. A complete
investigation would require renormalization group analysis, but
this is beyond the scope of this paper. We thus recovered our
universal behavior from before.

\section{Conclusions}
\label{viisi}

When analyzing many physical situations, it is common practice to
isolate the order parameter field, and more generally, the light
degrees of freedom, since these are expected to be the relevant
states at low energies. While this procedure certainly is
reasonable, in nature most of the physical fields are neither
order parameter fields, nor light at all. In order to extract
information from these heavy states, we needed first to determine
new and universal features associated with them. We have used a
general strategy proposed first in
\cite{Sannino:2002wb,{Mocsy:2003tr}}, according to which we couple
the light degrees/order parameter fields to the heavy fields in
the most general way, and then truncate the theory by retaining
all the relevant and marginal operators in the Lagrangian. In
doing so, the theory is fully renormalizable while capturing the
relevant contributions. The operator set is further constrained by
imposing all the relevant symmetries of the problem at hand. In
order for our procedure to work, we also assume, that the other
physical states of the theory have masses larger than our
non-order parameter field. In this way we can, formally, integrate
these states out, and their effects are absorbed in the modified
couplings of our effective Lagrangian.

For both the time independent and time dependent order parameter
field we have shown, that the spatial correlators of the non-order
parameter field are infrared dominated, and hence can be used to
determine the properties of the phase transition. We have
determined the general behavior of the screening mass of a generic
singlet field, and have shown how to extract all the relevant
information from such a quantity. Somewhat surprisingly, we have
demonstrated, that the pole mass of any non-order parameter
physical field is not infrared dominated. Our results can be
immediately applied to any generic phase transition. We have used
as relevant example, for the time independent order parameter
field case, the deconfining transition of Yang-Mills theories as
also explained in \cite{Sannino:2002wb,{Mocsy:2003tr}}. Lattice
simulations \cite{Bacilieri:mj} for the deconfining phase
transition support our results. Different approaches have been
used in literature to study the deconfining phase transition
\cite{{Boyd:1996bx},Sannino:2002wb,Agasian:fn,Campbell:ak,Simonov:bc,
Sollfrank:du,Carter:1998ti,Pisarski:2001pe,KorthalsAltes:1999cp,
Dumitru:2001xa,Wirstam:2001ka,Laine:1999hh,Sannino:2002re,
{Agasian:2003yw}}. Here we have been able to unify some of them.
In fact, by demonstrating that there exists an extended universal
behavior for singlet fields, we have uncovered the relation
between phenomenologically oriented models which use the glueball
Lagrangian
\cite{Agasian:fn,{Campbell:ak},{Simonov:bc},{Carter:1998ti},{Sannino:2002re}}
to describe the deconfining phase transition, and the ones using
the symmetries of the Polyakov loops
\cite{{Svetitsky:1982gs},Pisarski:2001pe,{Dumitru:2001xa},
{KorthalsAltes:1999cp},{Scavenius:2001pa},{Scavenius:2002ru}}.

The induced critical behavior is universal, however the
quantitative details depend on the strength of the couplings
between the fields as well as the resummation procedure. We have
considered only one non-order parameter field, but many are
expected to display a similar behavior. For the Yang-Mills
deconfining phase transition lattice QCD simulations are able to
determine the coupling strength of any glueball state to the
Polyakov loop by following the temperature dependence of screening
masses of such states.

Our analysis suggests, that monitoring a number of spatial
correlators, or more specifically their derivatives, is an
efficient and sufficient way to experimentally uncover the
chiral/deconfining phase transition and its features.

\acknowledgments It is a pleasure to thank P.H.~Damgaard,
A.D.~Jackson, J.~Kapusta, C.~Marchetti, and J.~Schechter for
discussions and careful reading of the manuscript. We acknowledge
R.~Pisarski for insightful discussions. Comments by E.~Kolomeitsev
are appreciated. The work of F.S.~is supported by the Marie--Curie
fellowship under contract MCFI-2001-00181.

\end{document}